\documentstyle[12pt]{article}
\begin{document}

\centerline{\Large \bf On the Clebsch - Gordan coefficients for the}
\centerline{\Large \bf two - parameter quantum algebra $SU(2)_{p,q}$}
\vspace{2.cm}
\begin{center}
{{\large Stjepan Meljanac}\\
\vspace{0.5cm}
Rudjer Bo\v skovi\' c Institute,\\
Bijeni\v cka c.54, 41001 Zagreb,\\
Croatia\\
\vspace{0.5cm}
{\large Marijan Milekovi\' c}\\
\vspace{0.5cm}
Prirodoslovno - Matemati\v cki fakultet,\\
Department of Theoretical Physics,\\
Bijeni\v cka c. 32, 41000 Zagreb,\\
Croatia}
\end{center}
\vspace{1.cm}
\centerline{Classification number: 02.20.+b}

\vspace{2cm}
\centerline{\large \bf Abstract}
\vspace{1.cm}

We show that the Clebsch - Gordan coefficients for the $SU(2)_{p,q}$
 - algebra depend on a single parameter  Q = $\sqrt{pq}$ ,contrary
 to the explicit calculation of Smirnov and Wehrhahn [J.Phys.A 25 (
 1992),5563].

\newpage
\vspace{1cm}
Recently, the Clebsch - Gordan problem for the two - parameter quantum
algebra $SU(2)_{p,q}$ was analyzed [ Smirnov and Wehrhahn 1992 ].It was
claimed that the corresponding C.-G. coefficients do depend on the two
deforming parameters p and q.

In this Comment we show that the C.-G. coefficients depend effectively
only on one parameter  Q = $\sqrt{pq}$, and that $SU(2)_{p,q}$ is
isomorphic to $SU(2)_{Q}$ , both as algebras and Hopf co - algebras.Our
results are in agreement with [ Drinfeld 1989 ].

We recall the  $SU(2)_{p,q}$  algebra defined in [ Smirnov and Wehrhahn 1992 ]
  (p and q are real parameters) :

\begin{eqnarray}
[J_{0},J_{\pm}] & = & \pm{J_{\pm}}
\nonumber\\
{[J_{+},J_{-}]}_{p,q} & = & J_{+}J_{-} - pq^{-1} J_{-}J_{+} ={[2J_{0}]}_{p,q}
\nonumber\\
{[2J_{0}]}_{p,q} &=&  \frac{q^{2J_{0}}-p^{-2J_{0}}}{q-p^{-1}}
\nonumber\\
{(J_{0})}^{\dagger} &=& J_{0} \hspace{20mm}   {(J_{\pm})}^{\dagger} = J_{\mp}
\end{eqnarray}

The coproduct $\Delta$ is :

\begin{eqnarray}
\Delta (J_{\pm}) &=& J_{\pm} \otimes p^{-J_{0}} + q^{J_{0}} \otimes J_{\pm}
\nonumber\\
\Delta (J_{0}) &=& J_{0} \otimes 1 + 1 \otimes J_{0}
\end{eqnarray}
The finite dimensional unitary irreducible representation  (IRREP)  $D^{j}$
  of spin  j  contains the highest weight vector $\mid jj>$, satisfying

\begin{eqnarray}
J_{0} \mid jj> &=& j \mid jj>
\nonumber\\
J_{+} \mid jj> &=& 0
\nonumber\\
<jj \mid jj> &=& 1
\end{eqnarray}
The other orthonormalized states of IRREP  $D^{j}$ , $\mid jm>$ , with
-j $\leq$ m $\leq$ j, satisfy

\begin{eqnarray}
J_{+} \mid jm>_{p,q} &=& (pq^{-1})^{{\frac{1}{2}}(j-m-1)} \sqrt{[j-m]_{p,q}
[j+m+1]_{p,q}} \mid jm+1>_{p,q}
\nonumber\\
J_{-} \mid jm>_{p,q} &=& (pq^{-1})^{{\frac{1}{2}}(j-m)} \sqrt{[j+m]_{p,q}
[j-m+1]_{p,q}} \mid jm-1>_{p,q}
\nonumber\\
J_{0} \mid jm>_{p,q} &=& m \mid jm>_{p,q}
\end{eqnarray}
Now, we define the $SU(2)_{Q}$ algebra with three generators $(J_{\pm})_{Q}$
and $(J_{0})_{Q}$ :

\begin{eqnarray}
[(J_{0})_{Q} , (J_{\pm})_{Q} ] & = & \pm (J_{\pm})_{Q}
\nonumber\\
{[(J_{+})_{Q} , (J_{-})_{Q} ]} & = & {[ 2 (J_{0})_{Q} ]_{Q}}
\nonumber\\
{[n]_{Q}} = {\frac{Q^{n}-Q^{-n}}{Q - Q^{-1}}} &=& {( \frac{p}{q} )^{ \frac{1}
{2}(n-1)} [n]_{p,q}}
\end{eqnarray}
with the coproduct
\begin{eqnarray}
{\Delta ((J_{\pm})_{Q})} &=& {(J_{\pm})_{Q} \otimes Q^{-(J_{0})_{Q}} +
Q^{+(J_{0})_{Q}} \otimes (J_{\pm})_{Q} }
\nonumber\\
{\Delta ((J_{0})_{Q})} &=& {(J_{0})_{Q} \otimes 1 + 1 \otimes (J_{0})_{Q} }
\end{eqnarray}
The relations between the $SU(2)_{p,q}$ generators and the $SU(2)_{Q}$
generators are
\begin{eqnarray}
J_{+} &=& {( \frac{q}{p})^{\frac{1}{2}(J_{0}- \frac{1}{2})} (J_{+})_{Q} }
\nonumber\\
J_{-} &=& {( \frac{q}{p})^{\frac{1}{2}(J_{0}+ \frac{1}{2})} (J_{-})_{Q} }
\nonumber\\
J_{0} &=& (J_{0})_{Q}
\end{eqnarray}
It is easy to show that relations  (7)  map equations (1) and equations (5)
 one  into another.Moreover, the $SU(2)_{p,q}$ coproduct is identical to the
  $SU(2)_{Q}$ coproduct, $\Delta_{p,q} \equiv \Delta_{Q}$:
\begin{eqnarray}
{\Delta_{p,q} (J_{0})} &=& {\Delta_{Q} (J_{0})} =
 {J_{0} \otimes 1 + 1 \otimes J_{0} }
\nonumber\\
{\Delta_{p,q} (J_{\pm})} &=& {\Delta_{Q} (J_{\pm})} =
\Delta_{Q}(( \frac{q}{p} )^{\frac{1}{2}(J_{0} \mp \frac{1}{2})} (J_{\pm})_{Q})
\nonumber\\
 &=& {J_{\pm} \otimes p^{-J_{0}} + q^{J_{0}} \otimes J_{\pm}}
\end{eqnarray}
This is also true for the antipode $\gamma_{p,q} \equiv \gamma_{Q}$ ,
the counit $\varepsilon_{p,q} \equiv \varepsilon_{Q}$ , the states \\
$\mid jm>_{p,q} \equiv \mid jm>_{Q}$ and the Casimir operator
$(C_{2})_{p,q} \equiv (C_{2})_{Q}$ .Thus we have proved the Hopf-algebra
isomorphism between $SU(2)_{p,q}$ and $SU(2)_{Q}$ .As a consequence
of this isomorphism, the p,q C.-G. coefficients of $SU(2)_{p,q}$ should be
identical to those of $SU(2)_{Q}$ .

Returning to the Smirnov and Wehrhahn's paper, one can immediately show,
using our equation (5) and $[n]_{p,q}= ( q/p)^{\frac{n-1}{2}}[n]_{Q}$
, that all the equations in Section 2 of their paper can be reduced to the
equations with single parameter Q.Particulary, the states in equation (2.7)
can be written as $\mid jm>_{p,q} = \mid jm>_{Q}.$ Therefore,the projection
operator $P^{j}_{mm'} = \mid jm><jm' \mid $ also has to be expressed in
terms of the Q - parameter only.However, their projection operator, eq.(3.7),
is wrong since it explicitly depends on the (p/q) parameter and does not
satisfy the projection property $(P^{j}_{mm})^{2} = P^{j}_{mm} $ .Hence,
all the subsequent formulae in Section 5 ( starting with the equation
(5.3)  ) are incorrect.For example, the C.-G. coefficients given in Table 1
do not satisfy the orthonormality relations (4.10).

The correct formula for the projection operator should read
\begin{eqnarray}
P^{j}_{mm'} & = & {( \frac{p}{q})^{- \frac{1}{4} (j-m)(j-m-1)} \hspace{2mm}
              \sqrt{ \frac{[j+m]_{p,q} !}{[2j]_{p,q}![j-m]_{p,q}!}}
    \hspace{2mm}J_{-}^{j-m} \hspace{2mm}P^{j} \hspace{2mm}J_{+}^{j-m'} }
 \nonumber\\
 & & {\times \hspace{1mm} ( \frac{p}{q})^{- \frac{1}{4} (j-m')(j-m'-1)}
              \sqrt{ \frac{[j+m']_{p,q} !}{[2j]_{p,q}![j-m']_{p,q}}}}
 \nonumber\\
 & \equiv & (P^j_{mm'})_{Q}
\end{eqnarray}
Using this projector we obtain the  C.-G. coefficients which depend on the
single parameter Q. In this way, these  C.-G. coefficients are identical
to those for  $SU(2)_Q :$
\begin{eqnarray}
<j_1 m_1 j_2 m_2 \mid JM>_{p,q}  =  <j_1 m_1 j_2 m_2 \mid JM>_{Q= \sqrt{pq}}
\end{eqnarray}
for all $j_1 , m_1 , j_2 , m_2 $ , J and M.

\newpage
{\large \bf Acknowledgments \bigskip}\\
This work was supported by the joint Croatian-American contract NSF JF 999
and the Scientific Fund of Republic of Croatia.

\vspace{2cm}
{\large \bf References \bigskip}\\
Drinfeld V G 1989 Algebra i analiz 1, 1 (in Russian)\\
Smirnov Yu F and Wehrhahn R F 1992 J.Phys.A:Math.Gen. 25  5563
\end{document}